\documentclass{aa}
\usepackage{graphicx}
\usepackage{epsfig}
\usepackage{subfigure}
\usepackage{txfonts}
\def\mso{\,\mathrm{M}_\odot}

 \newcommand{\gcm}{\,{\rm g}\,{\rm cm}^{-3}}
 
 \def\kms{\, {\rm km}\, {\rm s}^{-1}}

 \def\simle{\mathrel{\hbox{\rlap{\hbox{\lower4pt\hbox{$\sim$}}}\hbox{$<$}}}}
 \def\simgr{\mathrel{\hbox{\rlap{\hbox{\lower4pt\hbox{$\sim$}}}\hbox{$>$}}}}
 \def\msoy{\, \mso~{\rm yr}^{-1}}
\begin{document}
   \title{Constraints on gamma-ray burst and supernova progenitors through circumstellar absorption 
lines. (II):  Post-LBV Wolf-Rayet stars}

   \author{A. J. van Marle \inst{1,2}
          \and
          N. Langer \inst{1}
          \and
          G. Garc{\'i}a-Segura \inst{3}
          }

   \offprints{A. J. van Marle, {\tt marle@udel.edu}}

   \institute{Astronomical Institute, Utrecht University, 
              P.O,Box 80\,000, 3508 TA Utrecht, The Netherlands
        \and
              Bartol Research Institute, University of Delaware, 
              102 Sharp Laboratory, Newark, 19716, Delaware, USA
         \and
             Instituto de Astronom{\a'i}a-UNAM, 
             APDO Postal 877, Ensenada, 22800 Baja California, Mexico
             }

   \date{Received / Accepted }

   \abstract
% context heading (optional)
{Wolf-Rayet stars are thought to be the progenitors of Type~Ib/c supernovae and
of long gamma-ray bursts.}
% aims heading (mandatory)
{As shown by van Marle et al. (2005), circumstellar absorption lines
in early Type Ib/c supernova and gamma-ray burst afterglow spectra may reveal 
the progenitor evolution of the exploding Wolf-Rayet star. While the quoted paper deals with
Wolf-Rayet stars which evolved through a red supergiant stage, we investigate here the 
initially more massive Wolf-Rayet stars which are thought to evolve
through a Luminous Blue Variable (LBV) stage.} 
% methods heading (mandatory)
{We perform hydrodynamic simulations
of the evolution of the circumstellar medium around a 60$\mso$ star, from the main
sequence through the LBV and Wolf-Rayet stages, up to core collapse. 
We then compute
the column density of the circumstellar matter along rays from the central light source
to an observer at infinity,
as a function of radial velocity, time and angle.
This allows a comparison with the number and velocities, or blue-shifts, of absorption
components in the spectra of LBVs, Wolf-Rayet stars, Type~Ib/c supernovae and gamma-ray burst afterglows.}
% results heading (mandatory)
{Our simulation for the post-LBV stage shows the formation of various absorption components.
In particular, shells with velocities in the range of 100\, km\, s$^{-1}$ to 1200\, km\, s$^{-1}$ 
are formed at the beginning of 
the Wolf-Rayet stage, which are, however, rather short lived; they dissipate
on time scales shorter than $50\, 000\,$yr. As the LBV stage is thought to occur 
at the beginning of core helium burning, the remaining Wolf-Rayet life time
is expected to be one order of magnitude larger.   }
% conclusions heading (optional)
{When interpreting the absorption components in the afterglow spectrum
of GRB 021004 as circumstellar, it can be concluded that the progenitor
of this source did most likely not evolve through an LBV stage,
as no intermediate velocity absorption
components are predicted to prevail until core collapse. 
However, a close binary with a late common-envelope phase (Case~C) may produce a circumstellar medium that 
closely resembles the LBV to Wolf-Rayet evolution, but with a much shorter Wolf-Rayet period.
This scenario can not be ruled out.}

   
  \titlerunning{Absorption lines in GRB and SN spectra (II)}
  \authorrunning{van Marle et al.}
  \maketitle
%
%________________________________________________________________

\section{Introduction}
     Circumstellar nebulae are created by the interactions of stellar winds with the surrounding 
medium. 
     Typically, a fast wind sweeps up the interstellar gas or a preceding slower wind and forms a 
shell that travels outward. 
     This is the case for planetary nebulae, where the fast post-AGB wind sweeps up the slow AGB 
wind (Kwok \cite{Kwok}).
     Stars massive enough to form a Wolf-Rayet star
     ($M \simgr 25 \mso$) undergo a red to blue evolution in the Hertzsprung-Russell diagram 
(Meynet \& Maeder \cite{MM00}), analogous to that of post-AGB stars. 
     The radius of the star decreases, which causes the escape velocity at the 
surface and therefore the wind velocity to increase. 
     Stars below about 40$\mso$ are thought to evolve into a red supergiant before
     entering the  Wolf-Rayet stage (Meynet \& Maeder \cite{MM05}). For more massive stars,
     the main sequence is followed by a luminous blue variable 
(LBV) stage, which lasts only for a very short while ($\sim$~10\,000 years; 
      Langer et al. \cite{LA94}, Meynet \& Maeder \cite{MM05}).
     During the LBV stage the mass loss rate is extremely high ($\sim 10^{-3}~\msoy$), while wind 
velocities are of the order of 100...1000~$\kms$. 
     After the LBV stage, the star becomes a Wolf-Rayet star, with a lower mass loss rate, but 
higher wind velocity, which sweeps up the preceding wind.
     Such interactions create circumstellar nebulae around Wolf-Rayet stars. 
     These nebulae are temporary features, which will eventually dissipate into the surrounding 
medium. 
     This corresponds to the fact that circumstellar nebulae can only be observed around 
  a fraction of the Wolf-Rayet stars (Miller \& Chu \cite{MC93}).

     Wolf-Rayet stars end their evolution by the collapse of the massive iron core which
     forms in their center,
     which is thought to --- at least in a fraction of the events --- produce a violent
     stellar explosion which becomes visible as Type~Ib/c supernovae and (in some cases) long gamma-ray bursts (GRBs). 
     These events are a bright source of radiation in the UV to visible spectral range, 
     with photons being produced either in the 
     supernova photosphere (e.g., Doggett \& Branch \cite{DB85}),
     or through the interaction between a relativistic jet and the surrounding medium 
     as GRB afterglow (Ramirez-Ruiz et al. \cite{RGSP05}, Eldridge et al. \cite{EGDM06}, Nakar \& Granot \cite{NG06}).
     To reach us, this radiation has to pass through the circumstellar medium, which was shaped by the stellar wind during the evolution of the star. 
     Thereby, a part of the radiation is absorbed, creating absorption lines in the spectrum, which can be used to analyze the content of the circumstellar bubble as to density, composition and velocity structure.
 
     Blue-shifted absorption lines caused by the circumstellar medium should be visible in the spectrum of Type~Ib/c and Type~II supernovae. 
     Dopita et al. (\cite{DECS84}) found narrow P Cygni profiles seen in H$\mathrm{\alpha}$ and H$\mathrm{\delta}$ in the spectrum of the Type~IIL~\object{SN 1984E} indicating a Wolf-Rayet wind with a velocity of about 3000 \mbox{km} \mbox{s}$^{-1}$.
     A 350 $\kms$ absorption line in the spectrum of Type~IIL~\object{SN 1998S} can be attributed
to a moving shell (Bowen et al. \cite{BRMB00}; Fassia et al. \cite{Fetal01}).
    However, supernova spectra often contain a large number of photospheric absorption or P~Cygni lines
from the supernova 
itself, which makes the spectral analysis more difficult. Furthermore, supernovae are observed 
only at low redshift, which means that many of the most useful absorption lines are only observable in UV.
    One may hope that dedicated searches for circumstellar absorption lines in early
supernova spectra may reveal many more cases.    

     So far the best absorption spectrum in a GRB has been found in the afterglow of gamma-ray burst \object{GRB 021004}.
     It was proposed that the large number of absorption lines for \ion{C}{IV} and \ion{Si}{IV} visible in this spectrum can be explained as the result of circumstellar wind and shells (Schaefer 
et al. \cite{Setal03}; Mirabal et al. \cite{Metal03}; Fiore et al. \cite{Fietal05}, Starling et al. 
\cite{Setal05}, Lazzati et al. \cite{Letal06}).

     Van Marle et al. (\cite{MLG05a} and \cite{MLG05b}, from here on referred to as Paper~I) described the evolution of the circumstellar medium around a 40 $\mso$ star (see also Garc{\a'i}a-Segura et al. \cite{GLM96}), and a method to quantify the number, blueshifts and relative strengths of absorption components as function of time  produced by this circumstellar medium, and compared this to the observations of \object{GRB 021004}. 
     Here, we compute the circumstellar medium around a 60$\mso$ star, similar to Garc{\a'i}a-Segura et al. (\cite{GML96}). 
     While the 40$\mso$ star passed through the red supergiant phase before becoming a Wolf-Rayet star, the 60$\mso$ star evolves from the main sequence to the LBV stage and then moves on to become a Wolf-Rayet star. 
     Therefore, a different circumstellar absorption pattern is expected.

%__________________________________________________________________

   \begin{table*}
   \label{tab:windpar}
      \caption{
              Characteristic quantities for the evolution of the 60 $\mso$ star (Z=0.02), adopted as input for our hydrodynamic simulations.
              }
      \begin{tabular}{p{0.4\linewidth}rrrrr}
         \hline
         \noalign{\smallskip}
             Phase & End of phase [\mbox{yr}] & $\Delta$t [yr] & $\dot{m}$ [$\msoy$] & $V$ [$\kms$] & 
$n_{\mathrm{photon}}$ [\mbox{s}$^{-1}$] \\
         \noalign{\smallskip}   
         \hline
         \noalign{\smallskip}

   Main Sequence          & 3.4499$\times10^6$ & 3.4499$\times10^6$       & 3.51$\times10^{-6}$ & 2420 & 
7.12$\times10^{47}$ \\
   LBV                    & 3.4796$\times10^6$ & 2.9700$\times10^4$       & 4.71$\times10^{-4}$ & 492 & 
6.65$\times10^{47}$ \\
   Wolf-Rayet             & 3.8861$\times10^6$ & 4.0650$\times10^5$       & 6.41$\times10^{-5}$ & 2250 & 
6.22$\times10^{49}$ \\

         \noalign{\smallskip}
         \hline
      \end{tabular}
   \end{table*}

\section{Simulating the circumstellar bubble}
\label{sec-evolsim}
    The evolution of the circumstellar medium around a massive star can be divided into three stages, as the result of the evolutionary track of the star. 
    A 60 $\mso$ star begins as a main sequence star, develops into an LBV and finally becomes a Wolf-Rayet star. 
    This means that in the circumstellar medium the following interactions take place (cf. Garc{\a'i}a-Segura et al. \cite{GML96}): 
    
  \begin{enumerate}
  
    \item First: an interaction between the fast, low density main-sequence wind and the interstellar medium. 
          This interaction creates a moving shell, driven outward by the high thermal pressure of the shocked wind material.
	  Such an interaction can be described analytically (Weaver et al. \cite{WCM77}).\\
    \item In the next phase the slow, dense LBV wind hits the bubble created by the main-sequence wind, creating a new shell (the LBV shell), which moves into the main sequence bubble.\\
    \item Finally, the massive, high velocity Wolf-Rayet wind sweeps up the remnants of its predecessors. 
          The Wolf-Rayet wind drives a third shell outward, which overtakes the LBV shell. 
	  Both shells are destroyed by the collision. 
	  The remnants continue to move outward.\\

  \end{enumerate}
    In order to model the evolution of the circumstellar medium we use the same method as described in Paper~I and van Marle et al. (\cite{MLG06a}, \cite{MLG06b}). 
    We have divided the evolution of the star into three stages (main sequence, LBV and 
Wolf-Rayet), each with constant wind velocity, mass loss rate and number of ionizing photons per second. 
    The average mass loss rate follows from the evolution of the total mass of the star for each of 
these periods. 
    The average wind velocity is chosen so that the total kinetic energy output of the star 
remains the same as for a fully time dependent model. 
    The number of ionizing photons is calculated from the surface temperature of the star, using a 
Planck emission curve.
    As input model we used the 60 $\mso$ model with a metallicity of Z=0.02 that was calculated by 
Schaller et al. (\cite{SSMM92}). 
    The resulting parameters for the stellar wind and photon count are given in Table~\ref{tab:windpar}. 
    The density of the interstellar medium is set at $10^{-22.5}$ $\gcm$.  
    The effect of photo-ionization was included in the simulation by calculating the Str{\a"o}mgren 
radius along each radial grid line and correcting the temperature and mean particle weight within 
this radius as described by Garc{\a'i}a-Segura \& Franco (\cite{GF96}) and Garc{\a'i}a-Segura et 
al. (\cite{GLRF99}).

    The hydrodynamical simulations were done with the ZEUS 3D hydrodynamics code by Stone and 
Norman (\cite{SN92}).
    We simulate the main sequence only in 1D. 
    Because they are highly unstable, the LBV and Wolf-Rayet wind interactions have to be computed in 
2D, so we have taken the end result of the 1D simulation and mapped this onto a 2D grid, as was 
described in Garc{\a'i}a-Segura et al. (\cite{GML96}, \cite{GLM96}). 
    The transition from 1D to 2D has to be made before the start of the LBV phase in order to 
include the instabilities in the LBV shell. 
    This is different from the 40 $\mso$ case described in Paper I, where we could simulate most of 
the red supergiant phase in 1D.
    Since we want to follow the evolution of the circumstellar medium to the moment of supernova, 
we have calculated the whole circumstellar bubble in 2D, instead of using only the inner part as 
was done by Garc{\a'i}a-Segura et al. (\cite{GLM96}). 
    This method is similar to the one used in van Marle et al. (\cite{MLG04}) and was also 
described in Paper~I.\\        
    A similar calculation for a 60 $\mso$ star was done completely in 2D by Freyer et al. 
(\cite{FHY03}).
    Models of the circumstellar medium of stars in a large interval of masses and metallicities 
where produced by Eldridge et al. (\cite{EGDM06}).

   \begin{figure}
   \centering
   \resizebox{\hsize}{!}{\includegraphics[width=0.95\textwidth,angle=-90]{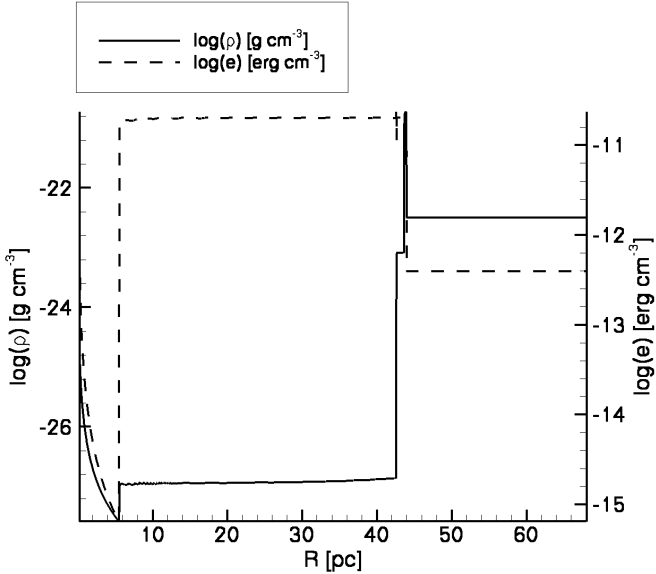}}
   \caption{Structure of the circumstellar bubble around our 60~$\mso$ star at the end of the main 
sequence phase (t~=~3.448 Myr). 
           This figure gives density (continuous line) and internal energy density (dashed line) as  
function of the radius. 
           From left to right we have: The freely expanding main sequence wind, the wind 
termination shock, the hot bubble with the small \ion{H}{II} region, the shell driven by the 
thermal pressure of the bubble and the interstellar medium. 
    }
   \label{fig:mainsequence}
   \end{figure} 
      
   \begin{figure}
   \centering
   \resizebox{\hsize}{!}{\includegraphics[width=0.95\textwidth]{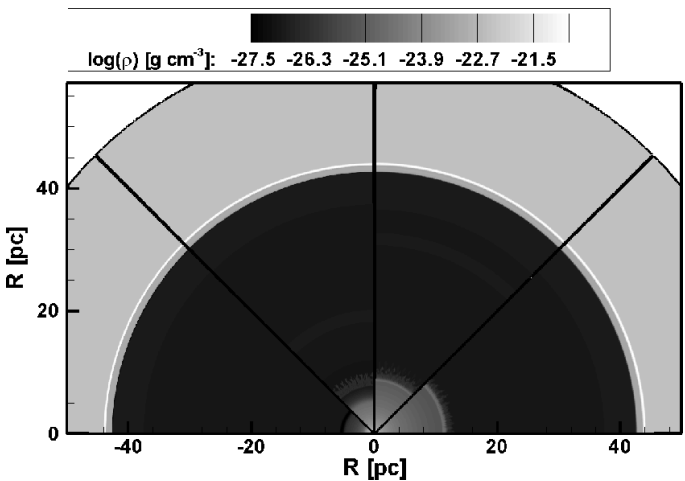}}
      \caption{The logarithm of the density [\mbox{g} \mbox{cm}$^{-3}$] of the circumstellar 
medium, during the LBV phase.
       Each segment shows a moment in time starting on the left at t~=~3.4516 Myr. 
       Each following segment is taken 7\,927 years later.
       The slow, dense LBV wind pushes a shell outward into the hot bubble. 
       The hot bubble itself pushes a second shell out into the interstellar medium. 
       }
         \label{fig:LBVbubble}
   \end{figure}

\section{The circumstellar bubble during main sequence and LBV phase}
    \label{sec-mstoLBV}
    \subsection{Main sequence phase}
    At the beginning of the main sequence phase the \ion{H}{II} region created by the radiation from a 
massive star pushes a shell into the interstellar gas. 
    At the same time the kinetic energy of the wind is converted into thermal energy by collision 
with the surrounding matter. 
    This creates a hot bubble which pushes a shell into the \ion{H}{II} region. 
    Unlike the case for the 40 $\mso$ star in Paper~I we have set the main sequence wind velocity at three 
time the escape velocity. 
    This, combined with the higher mass loss rate of the 60 M$_\odot$ star raises the thermal 
pressure in the bubble to the point where the shell driven into the \ion{H}{II} region moves 
supersonically. 
    Therefore, no pressure equilibrium between wind bubble and \ion{H}{II} region can be reached 
and the wind driven shell sweeps up the entire \ion{H}{II} region. 
    (In the case of the 40 $\mso$ model an \ion{H}{II} region can exist outside the wind bubble, 
even if the wind velocity is set at three times the escape velocity.)
    The end result is a hot bubble with nearly constant density, which drives a shell into the 
cold interstellar medium (Fig. \ref{fig:mainsequence}). 
    Close to the star is the free streaming wind, where the density and internal energy decrease 
with the square of the radius. 
    At at radius of ca.~6~pc the wind hits the termination shock where it slows down abruptly. 
    Its kinetic energy is converted into heat, causing a sudden increase in internal energy. 
    At 45 pc this hot bubble ends. 
    Here the internal energy is converted back into kinetic energy as the thermal pressure drives a 
shell outward. 
    The small density jump just before the shell is caused by the radiation from the star, which 
ionizes a small \ion{H}{II} region beyond the wind bubble. 
    This \ion{H}{II} region is very small, especially compared to a similar region found around 25...40 
$\mso$ stars (van Marle et al. \cite{MLG04}, \cite{MLG05a} and Paper~I).
    If the main sequence phase is modeled in two dimensions, the ionizing radiation will break through 
the wind driven shell, creating local 'fingers' of photo-ionized gas, which reach out from the wind 
driven bubble (Freyer et al. \cite{FHY03}).

    \subsection{LBV phase}
    During the LBV phase the mass loss rate increases dramatically while the wind velocity 
decreases
    As a result, a new shell is created once the LBV wind reaches the termination shock. 
    This shell, driven by the LBV wind, moves outward into the main sequence bubble. 
    At the end of the LBV phase, the circumstellar medium is built up as follows (see also 
Fig.~\ref{fig:LBVbubble}). 
    Closest to the star is the freely expanding LBV wind. 
    This area extends to ca.~12~\mbox{pc}.
    The wind termination shock itself is marked by a thin shell of shocked LBV wind material, which is driven 
outward by the wind. 
    Unlike the corresponding shell during the red supergiant phase of a 40 $\mso$ star 
(Garc{\a'i}a-Segura et al. \cite{GLM96}; van Marle et al. \cite{MLG04}, \cite{MLG05a}, 
Paper~I), the LBV wind driven shell has a significant radial velocity. 
    This is the result of the ram pressure of the LBV wind, which is several orders of magnitude 
higher than the ram pressure of the red supergiant wind. 
    Next comes the hot isotropic bubble, which in turn pushes a shell (r~$\sim$~45 pc in 
Fig.~\ref{fig:LBVbubble}) into the interstellar medium.

   \begin{figure}
   \centering
   \resizebox{\hsize}{!}{\includegraphics[width=0.95\textwidth]{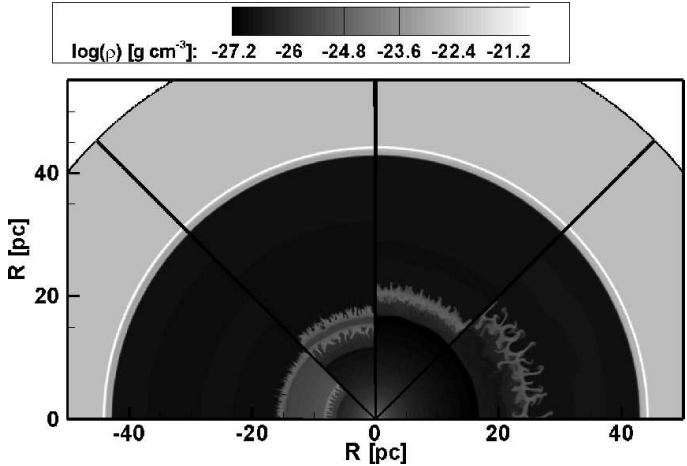}}
      \caption{Similar to Fig. \ref{fig:LBVbubble}, but during the early WR phase (transition from 
WR1 to WR2 in Section \ref{sec-WR}).
       Each segment shows a moment in time starting on the left at t~=~3.4881 Myr. 
       Each following segment is taken 7\,927 years later. 
       The fast Wolf-Rayet wind sweeps up the LBV wind in a shell, which overtakes the earlier LBV 
shell. 
       Both shells are fragmented during the collision and the fragments continue to travel 
outward. 
       Eventually, they will collide with the main sequence shell and dissipate into the 
hot bubble.        
       }
         \label{fig:WRbubble}
   \end{figure}

   \begin{figure}
   \centering
   \resizebox{\hsize}{!}{\includegraphics[width=0.95\textwidth]{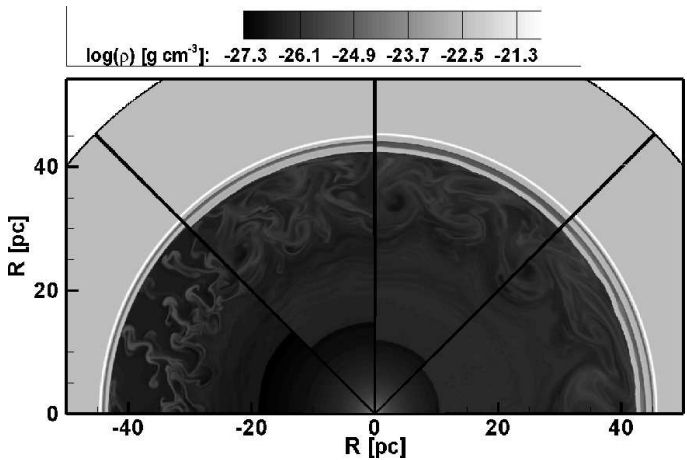}}
      \caption{Similar to Figs. \ref{fig:LBVbubble} and \ref{fig:WRbubble}, this figure shows the 
collision of the shell fragments with the outer edge of the hot bubble (transition from WR2 to WR3 
in Section \ref{sec-WR}). 
      The first segment (left) shows the density of the circumstellar medium at t~=~3.5436 Myr. 
      Since this process takes more time than the ones shown before, the time difference between 
segments has been increased to 47\,565 years
      The shell fragments travel outwards in the bubble and collide with the outer edge. 
      Afterwards, they dissipate into surrounding medium.
      The area between the low density bubble and the thin shell at r~$\simeq$~47 pc is the 
\ion{H}{II} region, created by high energy photons from the Wolf-Rayet star.
       }
         \label{fig:WRbubble2}
   \end{figure}

\section{The circumstellar bubble during the Wolf-Rayet phase}
    \label{sec-WR}
    The Wolf-Rayet phase can be divided into three separate parts (based on the evolution of the 
circumstellar medium, not the evolution of the star).

    During the first phase (WR1), starting at t~$\simeq$~3.48 Myr, the fast, high density WR wind 
drives a shell into the surrounding medium (the LBV wind remnant). 
    This shell moves rapidly outward, much faster than the shell driven by the LBV wind.

    The second phase (WR2) starts as the two shells collide (t~$\simeq$~3.5 Myr). 
    Both shells are highly unstable already and break up during the collision. 
    This process can be seen in Fig.~\ref{fig:WRbubble}, where we see the Wolf-Rayet wind driven 
shell overtake the LBV shell. 
    Unlike the collision between red supergiant and Wolf-Rayet shells as described in Paper~I, the remnants of the two shells are not completely fragmented, and there is less independent turbulent movement of the individual fragments. 
    They remain more or less together while they travel into the main sequence bubble. 
    This is the result of both the relative densities and velocities of the two shells 
upon collision. 
    The Wolf-Rayet shell is approximately ten time as dense as the LBV shell. 
    In contrast, the red supergiant shell has about the same density as the Wolf-Rayet shell 
(Paper I).
    The velocity difference is also much greater here, 
    as the Wolf-Rayet shell overtakes the LBV shell with a relative velocity of ca.~400 $\kms$, twice 
as much as the relative velocity of the Wolf-Rayet shell vs. the red supergiant shell in Paper I.
    As a result, the Wolf-Rayet shell is not completely destroyed and absorbs most of the LBV shell. 
    This phase lasts only for a short while, until the shell fragments hit the edge of the bubble. 

    The final part of the Wolf-Rayet phase (WR3) comprises the time after t~$\simeq$~3.53~Myr, which is when 
the shell fragments collide with the edge of the main sequence bubble (Fig.~\ref{fig:WRbubble2}).
    This phase last much longer than WR1 and WR2. 
    During this phase their is only one isotropic bubble, which is heated by the kinetic energy of 
the Wolf-Rayet wind and drives a shell into the interstellar medium. 
    The kinetic energy of the wind is high, which causes an increase in the temperature of the wind 
bubble, which in turn accelerates the movement of the shell. 
    Also, during the Wolf-Rayet phase the number of high energy photons increases, which results in 
a photo-ionized \ion{H}{II} region outside the wind bubble. 
    This means, that the hot bubble starts to show a more pronounced density discontinuity than 
during the main sequence, similar to the one we found for the 40 $\mso$ star (van Marle et al. 
\cite{MLG04}, \cite{MLG05a} and Paper I). 
    However, for the 60 $\mso$ star the width of the \ion{H}{II} region is very small ($\lesssim$~5~pc) 
compared to the total size of the hot bubble (Fig.~\ref{fig:WRbubble2}). 
    
   \begin{figure*}
   \centering
   \includegraphics[width=0.95\textwidth]{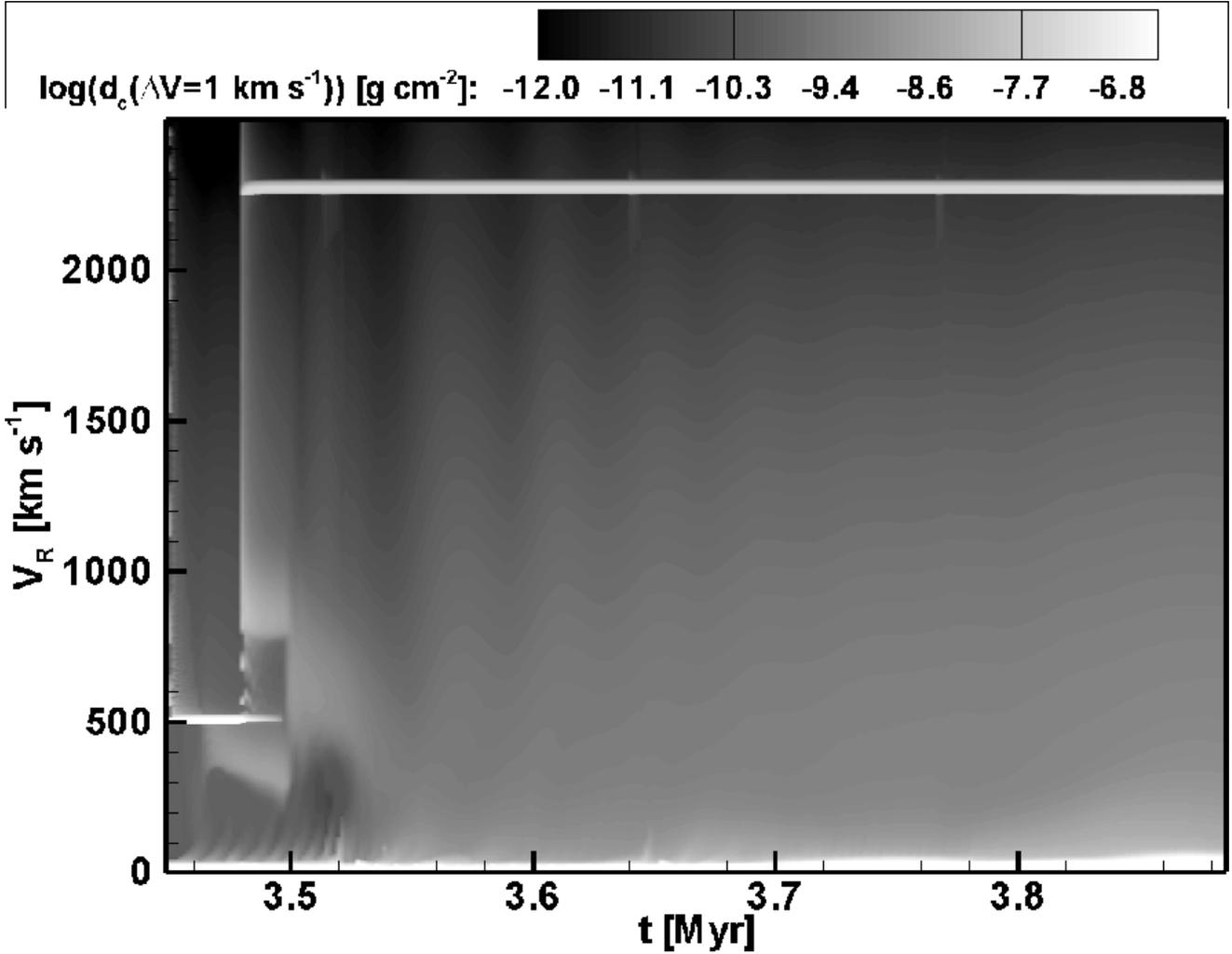}
      \caption{Column density $\mathrm{d}_{\mathrm{c}}(\varv,\Delta \varv)$, with $\Delta \varv$~=~1 
$\kms$, of the circumstellar medium during the LBV and Wolf-Rayet period as a function of radial 
velocity and time, averaged over 200 radial grid lines.
      On the horizontal axis is the time since the birth of the star in years. 
      The vertical axis displays the radial velocity in $\kms$.
      The plot starts at the end of the main sequence phase as the wind makes the transition from the fast main sequence wind to the slow LBV wind. 
      As the LBV phase starts a narrow feature appears for the LBV wind (500~$\kms$) and a broader feature for 
the LBV wind driven shell (300...400~$\kms$). 
      When the star becomes a Wolf-Rayet star, the Wolf-Rayet wind (2225...2275~$\kms$) and the shell (800...1000~$\kms$) driven by this 
wind both form independent features as well. 
      Once the Wolf-Rayet shell overtakes the LBV shell, both shells and the LBV wind disappear and 
a new shell is formed (500...800~$\kms$). 
      The fragments of this shell eventually collide with the outer edge of the bubble and 
dissipate. 
      Afterwards, only two independent features remain: the Wolf-Rayet wind and the 0...50~$\kms$ velocity 
component.
      }
         \label{fig:cd_all}
   \end{figure*}   

   \begin{figure*}
   \centering
   \includegraphics[width=0.95\textwidth]{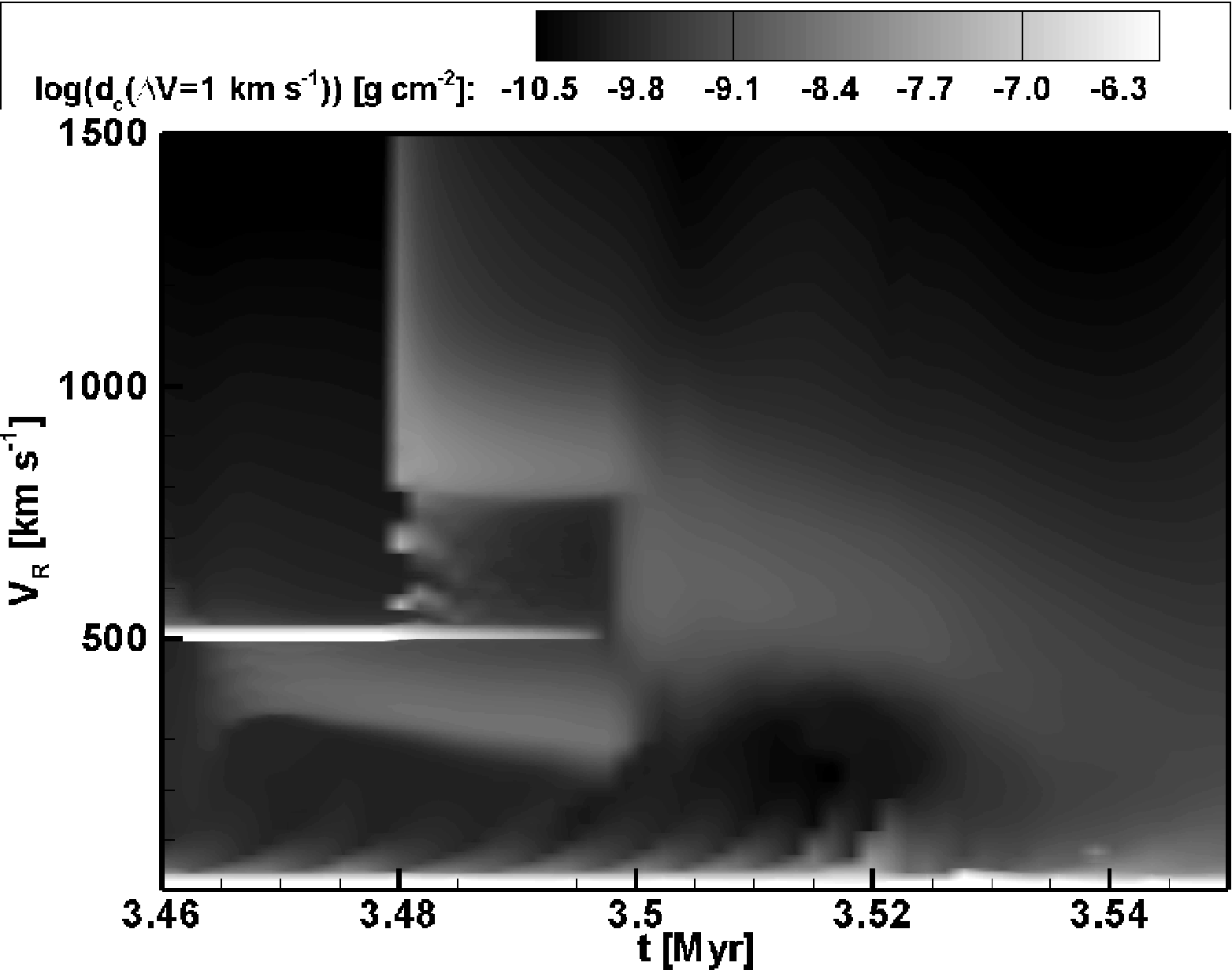}
      \caption{This figure shows a magnification of the most interesting part of Fig. \ref{fig:cd_all}. 
      It focuses on the time period between the start of the LBV phase and the end of phase WR2. 
      The first absorption feature to appear is the thin line at 490 $\kms$ created by the LBV 
wind. 
      Once this wind has reached the termination shock it forms a shell, which causes the broad, 
low velocity component (300...500~$\kms$). 
      Once the Wolf-Rayet wind starts, a new shell at ca.~800 $\kms$ is created (immediately, since 
this shell is created by sweeping up the preceding wind, rather than by collision with the wind 
termination shock). 
      This shell sweeps up the LBV wind, so the LBV wind feature disappears. 
      Once the Wolf-Rayet shell reaches the LBV shell, both shells disappear. 
      A new shell is created where the Wolf-Rayet wind meets the hot bubble. 
      This shell shows up as a new broad feature at ca.~600 $\kms$, which lasts until its 
fragments hit the edge of the bubble and dissipate.       
      }
         \label{fig:cd_detail}
   \end{figure*}

\section{Calculating the column density}
     \label{sec-abscalc}
     From our simulation of the circumstellar medium we calculate the column density as a function 
of radial velocity and angle for fixed times, as described in Paper I. 
     We move outward from the star along each radial grid line and take at each grid point the local 
column density (density multiplied with the radial length of the grid cell), the radial velocity 
(rounded to 1 $\kms$ intervals) and the temperature. 
     This procedure gives us the column density as a function of radial velocity. 
     We take thermal broadening into account by spreading the column density of each grid cell over 
a velocity interval that is calculated from the Maxwell-Boltzmann distribution for particle 
velocities at a given temperature. 
     This gives us the following quantity:
   \begin{equation}
   \label{equ:cdperinterval}
      \mathrm{d}_{\mathrm{c}}(\mathrm{v_r},\Delta \mathrm{v_r}) = 
\int_{\mathrm{r}=0}^{\mathrm{r}=\mathrm{R}} \rho (\mathrm{r}) \mathrm{P}(\mathrm{v}_\mathrm{r}, T) 
\Delta \mathrm{v}_\mathrm{r} \mathrm{dr} ,
   \end{equation}
     with: \mbox{v}$_r$ the radial velocity, $\rho$ the mass density and \mbox{P} the probability 
function for a particle to have a velocity in a given velocity interval along a single axis. 
     For the interval $\Delta$\mbox{v}$_r$ we use 1~$\kms$. 
     The outer limit of the integral, $R$, is the outer boundary of the hydrodynamical grid.
     The quantity d$_{\rm c}$ is the column density 
     per velocity interval.
     As we do not compute the chemical structure of our gas distribution, 
     true column densities of individual ion species can not be predicted accurately.

\section{Chemical composition of the circumstellar bubble}
     In our simulation we do not regard the chemical composition of the gas. 
     In order to get a quantitative analyzes of blue-shifted absorption features, one would 
need to know both the metallicity of the gas and the ionization states. 
     While we do not simulate these quantities we can say a few things about them. 
     The metallicity of the circumstellar bubble is of course a direct result of the composition of 
the stellar wind, which in turn depends on the evolutionary state of the star. 
     During the main sequence phase, the wind will be almost purely hydrogen.
     The composition of the Wolf-Rayet wind changes over time, from mostly helium with some 
hydrogen during the early Wolf-Rayet phase, to helium and carbon during the later stages. 

     The shell driven by the main sequence wind consists solely of interstellar matter. 
     The shell driven by the LBV wind into this bubble will be composed mostly of helium enriched LBV wind material 
and some matter of the hot bubble itself. 
     The third shell, driven by the Wolf-Rayet wind into the LBV wind, will at first consist only 
of LBV wind material that has been swept up. 
     After the collision with the LBV shell, the Wolf-Rayet shell will have direct contact with the 
hot bubble and sweep up some of the material in this bubble.
     The composition of the material in the bubble varies over time, depending on which winds have 
fed material into the bubble up to that moment but is dominated by hydrogen at all times.
     
     The ionization of the material depends on both the temperature and the number of ionizing 
photons. 
     Other than in the 40~$\mso$ case (see Paper I), the 60~$\mso$ star always produces a large 
number of highly energetic photons, which means that photo-ionization will always be important. 
     The hot bubble has a very high temperature (T~$\geq$~10$^6$ K).
     The shells are less hot, since their high density causes them to cool down efficiently through 
radiative energy loss. 
     After a supernova and especially after a gamma-ray burst, the degree of ionization will be 
even higher, since such events produce a massive amount of high energy radiation, which will pass 
through the circumstellar bubble.
(Prochaska et al. \cite{PCB06}, Chen et al. \cite{Cetal06} and 
Lazzati et al. \cite{Letal06})

\section{The absorption features during the LBV and Wolf-Rayet phase}         
     The result of the column density calculations described in Section \ref{sec-abscalc} can be seen in Fig. 
\ref{fig:cd_all}.
     Compared to the 40 $\mso$ case (Paper I) it is clear that a 60~$\mso$ star 
produces more time dependent features.
     
     \subsection{Number of visible absorption features}
     How many absorption features could be visible changes with time. 
     For the moment we shall ignore any angle dependence and 
look only at the evolution of the angle averaged features (Figs.~\ref{fig:cd_all} and \ref{fig:cd_detail}). 
     During the main sequence phase, only the zero velocity component  and the main sequence wind are visible. 
     While the main sequence shell has a certain radial velocity, this is too small to be observed 
independently.
     Once the LBV phase starts, two new features are created. 
     The LBV wind itself, which moves outward at a velocity of 490 $\kms$, and the shell driven by 
the LBV wind, moving at ca. 400~$\kms$. 
     It takes about $10^4$~years to create this shell, since the LBV wind has to travel out to the wind 
termination shock at $\sim 6$\,pc.

     As the star enters the Wolf-Rayet phase, another two features are added: 
     The Wolf-Rayet wind, moving at 2250~$\kms$ and the Wolf-Rayet wind driven shell, which moves 
at ca.~900~$\kms$. 
  The wind creates a thin feature ($\lesssim 50~\kms$), whereas the shell feature is about 200~$\kms$ wide.
     This means, that during the phase WR1 (see Sect.~\ref{sec-WR}), a total number of five independent 
absorption features is visible (Wolf-Rayet wind, the remnant of the LBV wind, two shells and the 
zero velocity component). 

     As the Wolf-Rayet shell expands outward, the area of free streaming LBV wind decreases, so the 
LBV wind absorption component starts to disappear. 
     Once the Wolf-Rayet shell reaches the LBV shell, the two shells collide and both disappear as 
independent features. 
     The remnants of the shell are fragmented and create a new shell, with a velocity of ca.~600 $\kms$. 
     This feature has a width of about 400~$\kms$.
     This shell exists throughout phase WR2, which therefore shows three absorption components: The 
Wolf-Rayet wind, the combined shell and the zero velocity component. 
     The period from the beginning of the LBV phase to the end of phase WR2, which shows the greatest 
time dependence, can be seen in more detail in Fig. \ref{fig:cd_detail}. 

     Once the shell fragments collide with the edge of the bubble, they dissipate and disappear as 
absorption feature. 
     the circumstellar medium now enters the final phase, WR3. 
     Only two absorption features remain, the Wolf-Rayet wind and the zero velocity component.

     \subsection{Line of sight dependence}
     Which and how many independent absorption features may be visible for a given line of sight 
depends on angle as well as on time.
     The narrow lines resulting from free-streaming winds are visible at any angle.
     The broader lines that result from the shells and shell fragments are angle dependent since 
the density of the shell is not constant. 

\begin{figure}
   \centering
   \resizebox{\hsize}{!}{\includegraphics[width=0.95\textwidth]{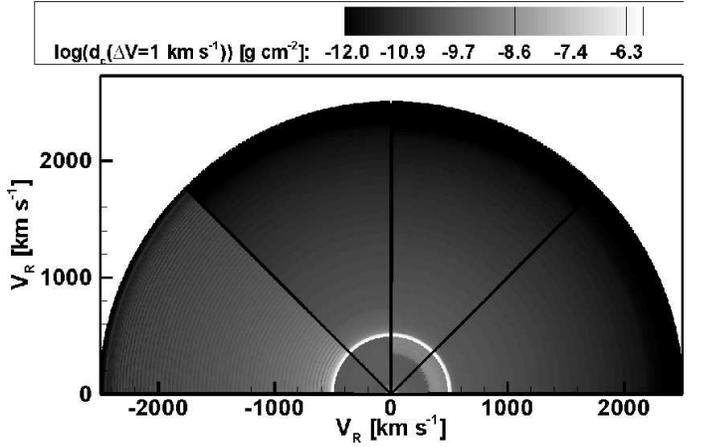}}
      \caption{The column density as a function of radial velocity and angle at the same moments in time as 
the density plot in figure \ref{fig:LBVbubble}.
      The LBV wind ($490~\kms$) is clearly visible at all times. The LBV shell ($\sim350-450~\kms$) 
appears later. Once it appears it can be observed at any angle.
      }
      \label{fig:angledepLBV}
   \end{figure}

\begin{figure}
   \centering
   \resizebox{\hsize}{!}{\includegraphics[width=0.95\textwidth]{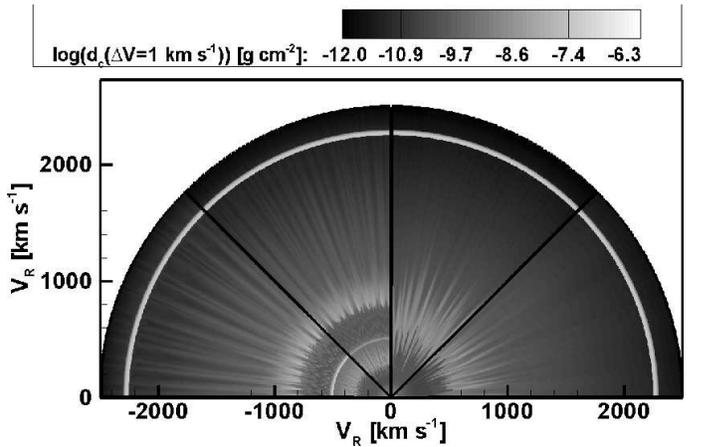}}
      \caption{Similar to Fig. \ref{fig:angledepLBV}, this shows the column densities at the same 
moments in time as the density plot in figure \ref{fig:WRbubble}.
       The Wolf-Rayet wind ($2250~\kms$) and the Wolf-Rayet shell ($800~\kms$) can be clearly 
observed in the first two frames  and the LBV wind and shell are still visible.
       The third and forth frame show the column densities after the collision of the Wolf-Rayet shell with the LBV shell. 
       The LBV wind and shell  and the Wolf-Rayet shell are no longer visible.
       A new shell moving at ca. 600~$\kms$ is formed from the fragments of the two shells and forms a new absorption feature.
       }
      \label{fig:angledepWR1}
   \end{figure}

\begin{figure}
   \centering
   \resizebox{\hsize}{!}{\includegraphics[width=0.95\textwidth]{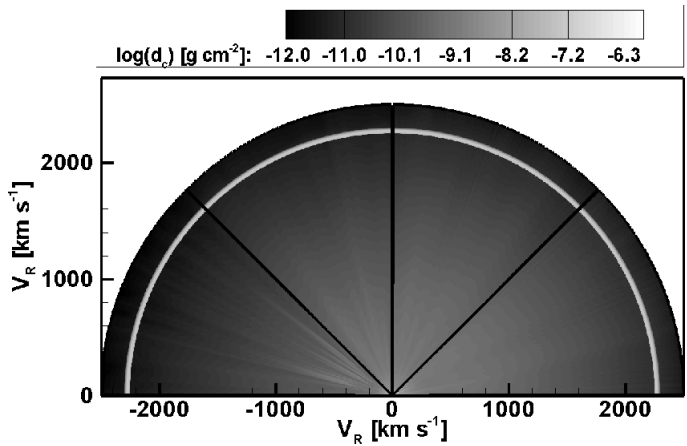}}
      \caption{The column density of the circumstellar medium at the same time as Fig. 
\ref{fig:WRbubble2}. 
      In the first frame, the remnants of the shells can still be seen as independent blobs (the 
column density varies considerably with the angle). 
      Once they have collided with the edge of the bubble, they slow down and spread out in angular 
direction.
      They will no longer show up as independent absorption features.
      }
      \label{fig:angledepWR2}
   \end{figure}

     In Figs.~\ref{fig:angledepLBV}, \ref{fig:angledepWR1} and \ref{fig:angledepWR2} we show the 
angle dependence of the absorption features during the LBV and Wolf-Rayet phase. 
     The segments shown correspond to the segments in Figs.~\ref{fig:LBVbubble}, \ref{fig:WRbubble} 
and \ref{fig:WRbubble2}.
     During the very early stages of the LBV phase, only the LBV wind itself is visible. 
     Once the LBV wind reaches the wind termination shock, it forms the LBV shell, which then 
becomes visible as an absorption feature (Fig.~\ref{fig:angledepLBV}).
     The LBV shell ($\sim 300...500~\kms$), though unstable, is not fragmented too severely and can 
be observed at any angle. 

     The start of the Wolf-Rayet phase can be seen in Fig. \ref{fig:angledepWR1}, which shows the 
column density during the transition from phase WR1 to phase WR2 (Section~\ref{sec-WR}). 
     The Wolf-Rayet shell ($\sim900~\kms$) becomes visible as an absorption feature as soon as the 
Wolf-Rayet wind ($2250~\kms$) starts.
     The WR shell is somewhat more fragmented, so the likelihood of observing this shell is not one, but still quite large.
     After the collision the fragments of the two shells form a new shell at ca.~600~$\kms$. 
     The angle dependence increases considerably after the collision of the two shells, as the 
fragments tend to shrink. 
     (The higher the local density, the more effective the radiative cooling. 
     Therefore, the high density fragments will have a low thermal pressure, which causes them to 
shrink even further.)
     As we saw in Paper I, fragmentation of a shell can both decrease and increase the number of 
independent velocity features seen in a single line of sight. 
     If no shell fragment is present along that line, no absorption line is visible. 
     However, it is also possible, that several fragments, with different individual velocities are 
present along the same line, causing multiple absorption features to appear. 
     The angle dependence plays a smaller role in the case of our 60~$\mso$ star than 
in the case of the 40~$\mso$ star of Paper I, since the shells around the more massive star are less fragmented.
     
     Once the shell fragments collide with the edge of the hot bubble, they cannot travel further 
outward and will instead spread out in the angular direction, as can be seen in 
Fig.~\ref{fig:angledepWR2}. 
     Since their radial velocity decreases, they will no longer show up as independent lines. 
     During the final stages of the Wolf-Rayet phase (WR3) only the Wolf-Rayet wind itself and the 
zero velocity component remain visible.

\section{Conclusions}
Our results show that the absorption spectrum created by the circumstellar medium around
a post-LBV Wolf-Rayet star is expected to be highly time-dependent. 
As in the case of a post-RSG Wolf-Rayet star (Paper~I), a fragmented intermediate velocity component
is formed by the snowplow effect of the Wolf-Rayet wind. 
Here however, the velocity of this component ($\sim 600\kms$) is somewhat higher 
than in the post-RSG case ($\sim 200\kms$) since the swept-up LBV shell itself
has already an appreciable velocity ($\sim 400\kms$). 
In practice, the two cases may be difficult to distinguish, however, as
shell fragments may have a considerably larger or smaller velocity than the
bulk of the shell (cf. Figs.~9 and~10 in Paper~I), and --- even though it is likely ---
 due to the clumpy structure
of the shell there is no guarantee that a given line of sight strikes a clump
which moves with the bulk velocity (cf. Figs.~8 and~9). 

    In comparison with the absorption lines observed in the afterglow spectrum of 
\object{GRB 021004}, the velocities of the absorption features predicted by
our 60$\mso$ model seem rather large. 
    However, there is an uncertainty due to the fact that the velocity of the LBV wind is 
difficult to predict. 
    We have adopted the escape velocity at the surface of the star as terminal wind velocity of the LBV wind,
     which is reasonable for a line driven wind. 
    However, the mechanism that drives LBV winds is not entirely clear and it is quite possible 
that the wind velocity is considerably lower. 
    (Garc{\a'i}a-Segura et al. \cite{GML96} found velocities as low as 200 $\kms$ for the LBV 
wind.)
    If the LBV wind has a lower velocity, so does the LBV shell which is driven by this wind. 
    The Wolf-Rayet shell, which has to sweep up the LBV wind, will be similarly affected.
    
The intermediate velocity (300...1000~$\kms$) absorption features only appear during the early 
Wolf-Rayet period and disappear approximately 50\,000 years after the start of the Wolf-Rayet 
phase, similar to the post-RSG case described in Paper~I.
    Since absorption lines appear at these velocities in the afterglow of \object{GRB 021004}, this 
would seem to indicate that the explosion took place during the early Wolf-Rayet stage.     
    However, this seems not possible for a post-LBV single star, as such stars  
have a Wolf-Rayet phase of at least 300\,000 years (Meynet \& Maeder \cite{MM05}). 
    A lower mass ($\sim 30~\mso$) star, as described in Paper I would seem to be a more likely 
candidate. 
    The large number of absorption features at intermediate velocities would then be explained by 
the presence of several shell fragments in the line of sight rather than as large scale evolutionary features.

    A second possibility is that the progenitor of \object{GRB 021004} was in fact part of a close 
binary, which went through a late mass-transfer phase. 
    During mass transfer, which last approximately 10\,000 years, the stellar wind parameters 
resemble those of the wind of an LBV star. 
    The mass donor star is stripped of its outer layers, so that only a Wolf-Rayet star remains 
(Petrovic et al. \cite{PLH05}). For so called Case~C mass transfer,
    this star will only have a short Wolf-Rayet phase left. 
    It may even be so short, that the wind and shell that were generated during the mass-transfer  
phase are still visible when the supernova occurs, which could account for the presence of a large 
number of intermediate velocity absorption lines without the need of multiple fragments of the same 
shell within the line of sight.     
 
     The assumption that stars of 40 $\mso$ or less can be gamma-ray burst progenitors is confirmed 
by observations of those gamma-ray bursts that can be linked to supernova explosions. 
     For \object{SN 1998bw}, which is identified as the supernova corresponding to  \object{GRB 
980425}, the initial mass of the star is considered to be $\leq$~40~$\mso$ (Iwamoto et  al. 
\cite{Ietal98}; Woosley et al. \cite{Wetal99} and Nakamura et al. \cite{Netal01}). 
     Since \object{GRB 980425} is not a typical gamma-ray burst, this can not serve as a general 
indication, but Mazzali et al. (\cite{Mazetal03}) give a similar result (30...40~$\mso$) for 
\object{SN 2003dh}, which corresponds to \object{GRB 030329}. 
     Finally, \object{SN 2003lw}, which is linked to to \object{GRB 031203}, is thought to have a 
progenitor mass on the main sequence of 40...50~$\mso$ (Mazzali et al. \cite{Mazetal05}). 
     This would make the star somewhat more massive than we expect, but it could still follow an 
evolutionary path similar to our 40~$\mso$ star model, rather than becoming an LBV star,
in particular if the metallicity of the star was sub-solar (cf. Langer \& Norman \cite{LN06}).
     In other words, all gamma-ray bursts that have been linked to supernova explosions so far 
seem to have a progenitor mass on the main sequence of less than 50~$\mso$ (see also della Valle 
\cite{DV05}).
     If the progenitor mass of \object{GRB 021004} is larger, the presence of the intermediate 
velocity lines in the afterglow spectrum can only be explained as the result of binary evolution.

  Of course, all of the above only pertains to a situation where the observed absorption features in the afterglow spectrum of \object{GRB 021004} are circumstellar in origin. 
  If their origin is interstellar (Prochaska et al. \cite{PCB06}, Chen et al. \cite{Cetal06}) it is not possible to deduce the evolution of the progenitor star from the absorption features.

\begin{acknowledgements} 
      We would like to thank Ralph Wijers, Rhaana Starling, Klaas Wiersema and Alexander van der 
Horst of the Astronomical Institute Anton Pannekoek of the University of Amsterdam for their 
information on the spectrum of the afterglow of \object{GRB 021004}.    
      This work was sponsored by the Stichting Nationale Computerfaciliteiten (National Computing 
Facilities Foundation, NCF), with financial support from the Nederlandse Organisatie voor 
Wetenschappelijk Onderzoek (Netherlands Organization for Scientific research, NWO).
      This research was done as part of the AstroHydro3D project:
      (http://www.strw.leidenuniv.nl/AstroHydro3D/)
\end{acknowledgements}

\end{document}